\documentclass[a4paper,12pt]{article}
\usepackage[]{graphicx}
\begin{document}
\title{Room-temperature stimulated emission from ZnO multiple quantum wells grown on lattice-matched substrates}
\author{Takayuki MAKINO, Yusaburo SEGAWA, \\
Masashi KAWASAKI$^{a)}$ and Hideomi KOINUMA$^{b)}$.}
\maketitle 
\begin{abstract}
The optical properties of ZnO quantum wells, which have
potential application of short-wavelength semiconductor laser utilizing a high-density excitonic effect, were investigated. Stimulated emission of excitons was
observed at temperatures well above room temperature due to the adoption of the lattice-matched substrates. The mechanism of stimulated emission from ZnO quantum wells is discussed in this chapter.
\end{abstract}

{\obeylines 
affiliations: 
Photodynamics Research Center, RIKEN (Institute of Physical and Chemical Research)
(519-1399, Aramaki aza Aoba, Aobaku, Sendai 980-0845)
E-mail:  tmakino@postman.riken.go.jp
a) Institute for Materials Research, Tohoku University,
(Sendai 980-8577, Japan)
b) Frontier Collaboration Research Center, Tokyo Institute of Technology
(4259 Nagatsuda, Midoriku, Yokohama 226-8503)}
\newpage
\section {Introduction --- ZnO as an optoelectronics material}

There have been many studies recently on the properties of
widegap semiconductors aimed at the development of a short-wavelength laser diode.
Akasaki and Nakamura developed a blue light-emitting diode and a continuous-wave laser in which InN--GaN alloys play an essential role as the active layers~\cite{akasaki1,nakamura_spr}.
However, in all commercially available semiconductor lasers including GaN-lasers, a recombination of electrons and holes is used as the
mechanism of laser action.
In such cases, the threshold carrier density required to accomplish the inversion distribution of a population for an electron--hole system is one or two orders of magnitude higher than the
Mott transition density.
If an exciton-related recombination is used as the mechanism of laser action, the resultant threshold value is expected to be two or three orders of magnitude lower with higher quantum efficiency.
Because of this expectation, interest has recently been shown in zinc oxide (ZnO), the band
gap of which is in the ultraviolet range.
In addition, the exciton binding energy (EBE) of ZnO is rather large
(60~meV) compared with those of other semiconductors such as group II-selenides
or group III-nitrides. Thus, the advantage of ZnO is that excitons can exist stably even at room temperature (RT) and even under a high-density condition.

Atomic layer control technology in the laser-assisted molecular-beam-epitaxy (LMBE) of oxides has advanced remarkably since the discovery of high-temperature superconductors.
In response to this, we started study of ZnO epitaxial growth for
the development of the short-wavelength semiconductor optoelectronics.
At first, we adopted sapphire as substrates that had large lattice-mismatch (18\%) with ZnO. Optical and structural properties of ZnO epitaxial layers
deposited on sapphire has been investigated.
These studies revealed that the ZnO epilayers are adapted for
the optoelectronic applications with the following aspects:
\begin{enumerate}
	\item Laser oscillation of excitonic origin was observed at room-temperature. Grain boundaries in the epilayers act as longitudinal cavities.
	\item Band gap energies can be tuned by preparing the (Zn,Mg)O and (Zn,Cd)O solid solutions~\cite{ohtomo1,kawasaki2,makino14}.
	\item Growth of ZnO/(Mg,Zn)O multi-quantum wells (MQWs) were succeeded.
	\item Optical properties of these MQWs were investigated. Excitonic luminescence accompanied by the quantum size effect at low temperatures was observed~\cite{ohtomo4,ohtomo7,makino17}.
\end{enumerate}

However, the abovementioned studies have also revealed some problems (unsatisfactory
properties of ZnO epilayers) that are unavoidable as long as lattice-mismatched substrates are used.
These ZnO thin films
are indeed epitaxial but stay multicrystalline in nature having
incoherent grain boundaries. These grain boundaries
seem to be useful for observing the
excitonic laser action because they can function
as mirrors of a longitudinal
cavity. The electrical properties of such films are, however,
rather poor, as represented by a typical electron concentration of
$n \sim 10^{17}$~cm$^3$ and a typical Hall mobility of $\mu \sim $10 cm$^2$/V s at room temperature. The electrical properties of the epilayers
are clearly inferior to those of bulk single crystals ($n \sim 10^{15}$~cm$^3$ and $\mu \sim $200 cm$^2$/V s). The crystallinity of epilayers
is also inferior.
In addition, the fact that neither luminescence (photoluminescence) nor a stimulated emission could be observed at room temperature in MQWs grown on the sapphire substrates is a problem.
The current advanced technology of compound semiconductors
cannot be fully utilized in the case of such inferior thin films and quantum wells (QWs).
Thus, QWs of higher quality must be developed to produce laser-diodes that can oscillate with a lower threshold. Furthermore, \textit{n}- and \textit{p}-type layers that have both high mobility and low resistance must be formed in order to perform current injection efficiently.

These problems might be resolved by using lattice-matched
substrates. We have adopted hexagonal
ScAlMgO$_4$(0001) with lattice constants of $a=3.246$~\textrm{\AA} and
$c=25.195$~\textrm{\AA} (Ref.~\cite{kimizuka1}) that have an in-plane lattice mismatch as
small as 0.09\%. ScAlMgO$_4$ is regarded as a natural
superlattice composed of alternating stacking layers of
wurtzite-type MgAlO$_x$ and rocksalt (111)-ScO$_y$
layers and hence has a cleavage habit along the (0001)
plane. High-quality single crystals can be grown by Czochralski's
method. The structure of a crystal grown by this method and a possible hetero-interface
with ZnO are schematically shown in Fig.~1 of Ref.~\cite{ohtomo5}.

Although
there is now a fairly good understanding of the basic properties of ZnO
epilayers, it is
only recently that the basic properties of its QWs have been studied in detail.
For example, Vispute \textit{et~al.} reported the growth and optical properties of
GaN/ZnO/GaN double heterostructures.
Although the combination of ZnO and (Zn,Mg)O has been used in a study by
another research group (Chen \textit{et~al.}~\cite{yfchen3}), more comprehensive
studies on this interesting and unexplored material have been conducted by us.
In this paper, we describe the optical properties of ZnO/(Zn,Mg)O MQWs grown
on lattice-matched ScAlMgO$_4$ (SCAM) substrates in detail.
This chapter is organized as follows. The experimental procedures are briefly described in Sect.~\ref{Sect:2}. Improvements in various properties of ZnO thin films achieved by the use of the lattice-matched substrates are described in Sect.~\ref{Sect:3}. In Sects.~\ref{Sect:4}--\ref{Sect:7}, quantum confinement effects of excitons, the well-width dependence of
the exciton-phonon coupling constants, and possible mechanism of a stimulated emission of ZnO MQWs are described respectively.
The summarizing remarks are given in Sec.~\ref{Sect:8}.
\section{Method for growing MQW samples}
\label{Sect:2}
Samples of MQWs (10 periods) were grown by the method of laser molecular-beam epitaxy. A QW is defined as stacks alternatively deposited by using two kinds of very thin semiconductor layers (wells and barriers) that have different
band gap energies. ZnO was used as a well layer material and a ZnMgO
solid solution, the band gap of which is larger than that of ZnO, was used as a barrier layer. The Mg concentration dependence of the band gap energy is given elsewhere~\cite{matsumoto1}. It should be noted that the in-plane
lattice mismatch between ZnO and these alloys is very small. The Mg concentration was set to 0.12 or 0.27, because the barrier height could be changed by a change in its
concentration. Eighteen samples with different Mg concentrations
and well widths ($L_w$), 6.9 to 46.5~\textrm{\AA}, were prepared in order to
estimate the $L_w$ dependence of their optical properties. Barrier layer thickness was fixed at 50~\textrm{\AA}. The samples were grown by the ``combinatorial'' method, the concept of which has been explained in the review article~\cite{koinuma2}. Readers should refer to related papers in which the apparatuses developed for efficient structural and optical
characterizations are described~\cite{ohtomo9,makino9,ohtani1,ohtani2}.

\section{Improvements in properties of ZnO thin films
achieved by the use of lattice-matched substrates}
\label{Sect:3}
Here, we describe briefly the improvements in the properties of ZnO epilayers achieved by using lattice-matched SCAM substrates. Observation of the surface morphology of ZnO/SCAM epilayers revealed that the steps have a flat terrace in the scale of an atomic level and have a height of 0.26~nm (corresponding to the charge neutral unit). Comparison of the full width at half maximum (FWHM) of the x-ray diffraction (XRD) curves showed that the crystallinity of these epilayers had reached almost the same level as that of a single crystal~\cite{ohtomo5}. These films also have high carrier mobility (100~cm$^2$/Vs) and low residual carrier concentration ($10^{15}$~cm$^3$). Even if the growth temperature is lowered to about 200~$^{\circ} $C, the high level of crystallinity was not changed. Recently, the formation of epilayers that have high resistance in which Hall measurement cannot be performed, and that have the high mobility ($\approx $200~cm$^2$/Vs)
has been achieved. In order to examine the activation efficiency of donor doping, \textit{n}-type samples were prepared by doping aluminum. The activation efficiency was about 10 times greater than that of a thin film grown on a sapphire substrate.
The optical properties of \textit{n}-type ZnO:Al has been reported elsewhere~\cite{makino19}.
\begin{figure}
\includegraphics[width=.5\textwidth]{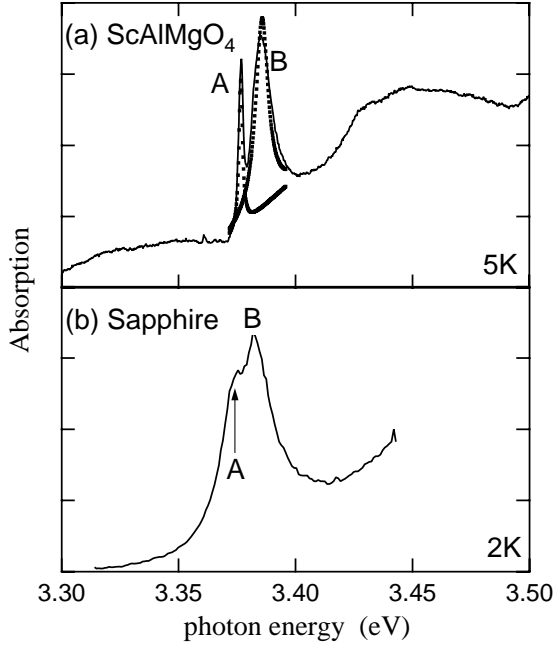}
\caption{Comparison of low-temperature absorption spectra
obtained from a ZnO epilayer grown on a sapphire substrate and that grown
on a SCAM substrate~\protect\cite{makino8}. Temperatures are shown on the
right-hand side of the figure. SCAM is an abbreviation of
ScAlMgO$_4$. ``\textit{A}, \textit{B}'' indicates \textit{A}- and \textit{B}-exciton absorption bands, and ``\textit{I}$_{6}$'' shows PL of a bound exciton state.}
\label{abs_epi}
\end{figure}
In ZnO, the \textit{A}-\textit{B} exciton splitting is only 7 or 8~meV~\cite{LBZincoxide}. These two excitonic peaks were clearly resolved in
the absorption spectrum measured at 4.2~K for bulk crystals~\cite{liang1}.
On the other hand, this was not the case for ZnO epilayers grown on
sapphire substrates~\cite{makino8} as shown in Fig.~\ref{abs_epi}.
The two excitonic peaks became too
broad to be resolved because the damping constant of the excitons became
larger due to the inferior crystallinity, as mentioned earlier. Such
undesired broadening could be avoided in the case of ZnO epilayers grown
on SCAM substrates. The FWHMs of the exciton absorption bands of these
samples were similar to those of bulk crystals~\cite{liang1,makino8}. It
was confirmed that the threshold for the stimulated emission is also
significantly improved~\cite{ohtomophd}. Such improvements in electrical,
structural and optical properties would not be possible if a sapphire
substrate is used. It can be concluded that quality of these single-crystalline ZnO epilayers satisfies the stringent requirements for being
regarded as compound semiconductors.

\section{Quantum confinement effect of excitons in quantum wells}
\label{Sect:4}
As mentioned earlier, the MQWs showed the following drawbacks due to the formation of rough interfaces caused by the use of lattice-mismatched substrates: (1) controllability of layer thickness is not sufficient for quantum confinement effect to be elicited in the case of $L_w \le 15$~\textrm{$\AA$}, and (2) PL efficiency is not high enough to enable observation of the exciton emission at RT~\cite{ohtomo4,ohtomo2}. These problems must be overcome for an optoelectrical device to be operable at RT. These problems, which were unavoidable when the sapphire substrates were used, could be eliminated by using ScAlMgO$_4$ substrates~\cite{makino11}.

The XRD patterns of the MQWs showed Bragg diffraction peaks and clear intensity oscillations due to Laue patterns corresponding to the layer thickness. This indicates a high crystallinity and a high degree of thickness homogeneity. Furthermore, observation of the atomic force microscopy (AFM) images revealed that the surface of an MQW is composed of well-defined atomically flat terraces and steps corresponding to the charge neutral unit of ZnO. Therefore, the interface roughness in the heterostructure cannot be larger than 0.26 nm. We conclude that ZnO and MgZnO alloy layers grow in a two-dimensional growth mode on this substrate, resulting in the formation of a sharp hetero-interface between them.

\begin{figure}
\includegraphics[width=.5\textwidth]{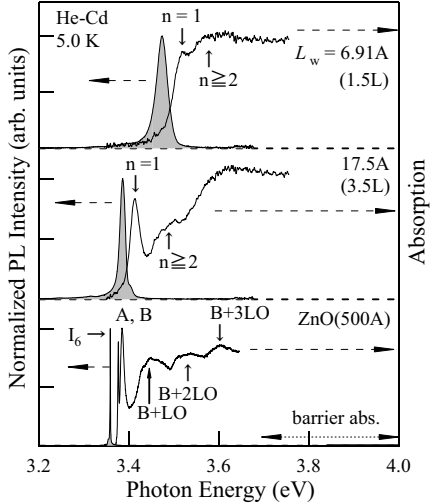}
\caption{PL and absorption spectra obtained from ZnO/Mg$_{0.12}$Zn$_{0.88}$O MQWs measured at 5~K for $L_{w} of 17.5$ and of 6.9~${\rm \AA} $. Absorption energy of barrier layers is shown by a horizontal arrow. Spectra obtained from a 500-${\rm \AA} $-thick ZnO film are also shown. ``B+LO, B+2LO, and B+3LO'' correspond to exciton-phonon complex transitions, ``$n = 1$'' shows the lowest excitonic absorption of the well layers, and ``$n \ge 2$'' means the excited states of the exciton or higher interband transitions.}	\label{abs_SL}
\end{figure}
Figure~\ref{abs_SL} shows PL and absorption spectra in ZnO/Mg$_{0.12}$Zn$_{0.88}$O MQWs on SCAM substrates measured at
5 K with well widths ($L_w$) of 17.5 and 6.9~\textrm{\AA}. The PL and absorption
spectra in a 500-\textrm{\AA}-thick ZnO epilayer on a SCAM substrate are included
for comparison. Both the PL and absorption peaks
shifted towards the higher energy side as $L_w$ decreased. This
shift was due to the quantum confinement effect. The exciton
Bohr radius is $\approx$18~\textrm{\AA}~\cite{LBZincoxide}. The absorption peaks ($n=1$) arise
from the lowest excitonic states of well layers. The peak
energies of PL were constantly located on the lower energy
side of those of absorption peaks.

\begin{figure}
\includegraphics[width=.5\textwidth]{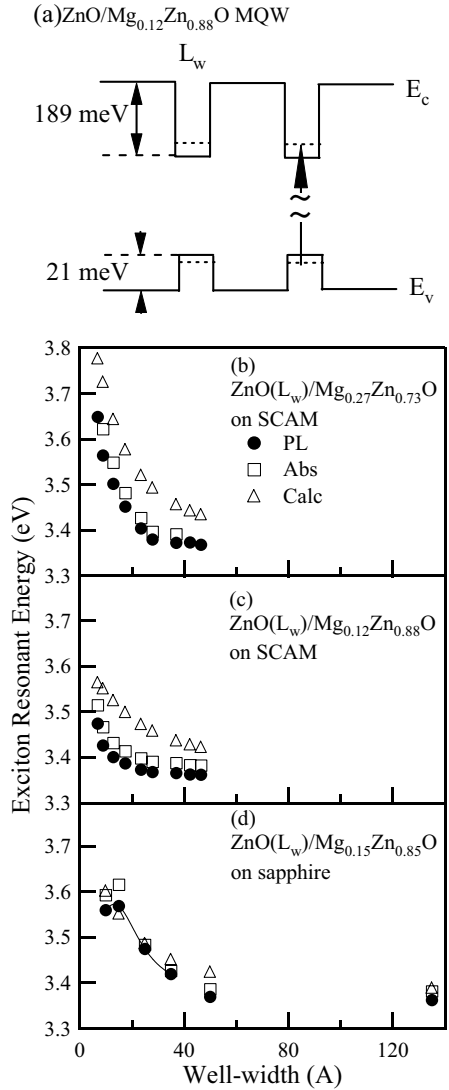}
\caption{(a) Diagram of conduction and valence bands between barrier and well layers in a ZnO/Mg$_{0.12}$Zn$_{0.88}$O MQW~\protect\cite{ohtomo4}. The upward arrow shows the lowest interband transition. (b) Peak energies of PL (circles) and absorption (squares) are plotted against $L_{w} $ in [ZnO/Mg$_{0.27}$Zn$_{0.73}$O]$_{10}$ on SCAM substrates. Results of calculation (triangles) of the interband transition energy that include the excitonic effect are also shown. (c) Similar except that the Mg content was $\approx 12$\%. (d) Similar except that the substrate was sapphire. The Mg content was $\approx 15$\%. Energies of PL excitation spectra (squares) are plotted instead of those of absorption spectra, due to the presence of 100-nm-thick ZnO buffer layers. Note that the peak energies of PL excitation spectra coincide with those of the absorption spectra. The curve is
 a visual guide~\protect\cite{makino11}.}
		\label{offset}
\end{figure}

Figures~\ref{offset}(b)--(c) show the well width dependence of the peak energies of PL (closed circles) and absorption (open squares), respectively~\cite{makino11}. The lowest transition energy of excitons (open triangles) formed with confined electrons and holes was calculated by using the model of one-dimensional, finite periodic square-well potential proposed by Gol'dman and Krivchnokov~\cite{goldman}. The EBE (59~meV, Ref.~\cite{huemmer1}) is assumed to be independent of $L_w$ here, although, as will be shown later, EBE actually depends on $L_w$.
The optical transition process on ZnO/Mg$_{0.12}$Zn$_{0.88}$O MQW is shown in Fig.~\ref{offset}(a). This tendency of the $L_w$ dependence of the exciton transition energy was qualitatively reproduced by calculation. As reported by Coli and Bajaj~\cite{coli1}, incorporation of the effects of
exciton-phonon interaction and quantum confinement in the calculation of the EBE, leads to the values of
the excitonic transitions that agree well with our experimental data.
Figure~\ref{offset}(d) shows the corresponding peak energy plot for MQWs grown on sapphire substrates. As seen in Fig.~\ref{fig:pkplot}(d), both of the peak energies have a maximum at $L_w$ of 15~\textrm{\AA} when sapphire substrates were used. This is a critical $L_w$ that prevents quantum confinement with respect to the exciton energy.
This is because of the poor controllability of $L_w$ due to the lattice mismatching.

\section{Well-width dependence of exciton-phonon interaction
in quantum wells}
\label{Sect:5}
\begin{figure}
\includegraphics[width=.5\textwidth]{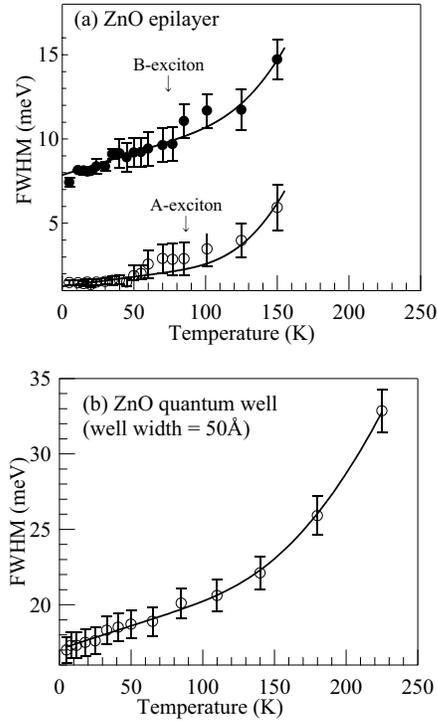}
\caption{(a) Width (full width at half maximum, circles) of \textit{A}- and \textit{B}-exciton absorption bands plotted as a function of temperature. Closed circles are data of the \textit{A}-excitons and the open circles are data of the \textit{B}-excitons. The solid curves represent the fitting results. (b) Similar plot for the MQW with Mg concentration of 0.12 and
$L_w$ of 46.5$\textrm{\AA }$~\protect\cite{makino8,sun2}.}
	\label{fwhm}
\end{figure}

The coupling constant between excitons and phonons in ZnO MQWs has not been estimated quantitatively. We therefore tried to quantify the coupling constant
by estimating the temperature dependence of the absorption spectra. Figure~\ref{fwhm} shows the temperature dependence of the full width at half maximum (FWHM) of the excitonic absorption peaks for a ZnO epilayer (a) and for a typical MQW sample with a QW width of $17.5$~$\textrm{\AA }$ (b). The solid line represents the fitted results based on the following equation. The temperature dependence of the FWHM can be approximately described by the following equation~\cite{lautenschlager1}:
\begin{equation}
\Gamma (T) = \Gamma _{0} + \gamma _{\rm ph} T+\Gamma _{\rm LO} /[\exp(\hbar \omega_{\rm LO}/k_{\rm B}T)-1], 
\label{e2}
\end{equation}
where $\Gamma _{\rm 0}$, $\hbar \omega_{\rm LO} $ (72~meV), $\gamma_{\rm ph} $, $\Gamma _{\rm LO} $ and $k_{\rm B}$ are the inhomogeneous
linewidth at temperature ($T$) of 0~K, longitudinal optical (LO)-phonon energy,
strengths of the exciton-acoustic-phonon and the exciton-LO-phonon couplings and the Bolzmann constant, respectively.
It was experimentally found that
$\hbar \omega_{\rm LO} $ of the MQWs is not different from the bulk value.

\begin{figure}
\includegraphics[width=.5\textwidth]{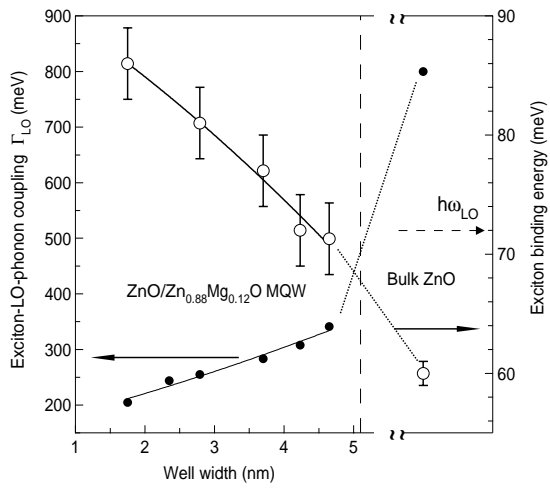}
\caption{Strengths of coupling between excitons and LO phonons $\Gamma_{LO}$ (closed
circles) and exciton binding energies (open circles) in bulk ZnO and
MQWs with different $L_w$s~\protect\cite{sun1}. The solid curve is a visual guide.}
\label{bindingenergy}
\end{figure}
Figure~\ref{bindingenergy} (closed circles, left axis) shows the values of $\Gamma _{\rm LO} $ obtained for the epilayer and its well-width ($L_w$) dependence obtained for ZnO/Mg$_{0.12}$Zn$_{0.88}$O MQWs. The values of $\Gamma _{\rm LO} $ of the MQWs are smaller than those for the epilayers and monotonically decrease with decrease in $L_w$. Here we try to explain this result by the enhancement of EBE induced by the quantum confinement effect. Figure~\ref{bindingenergy} (open circles, right axis) shows $L_w$
dependence of EBE. This dependence was determined
by studying spectra of stimulated emission. As is well known, the major process that contributes to broadening of the exciton linewidth is scattering of 1S excitons into free-electron-hole continuum or into excited excitonic states by absorbing LO phonons. If EBE exceeds $\hbar \omega_{\rm LO} $ (72~meV), dissociation efficiency into the continuum states is greatly suppressed compared to the
case of EBE smaller than $\hbar \omega_{\rm LO} $. In such case, $\Gamma _{\rm LO} $ is reduced. Indeed, EBE exceeds $\hbar \omega_{\rm LO} $ in the case of MQWs. A similar effect has also been
observed in other QW systems~\cite{pelekanos1}. Schematic explanation is described in detail in corresponding original papers~\cite{sun2,sunfull1}.

\section{The localization mechanism of the exciton in a quantum well}
\label{Sect:6}
It was found that the excitonic luminescence in the ZnO MQWs under investigation
is due to radiative recombination from excitons localized
by the potentials formed by the fluctuations of $L_w $ and barrier height.
Our spectral assignment are based on (1) the well width dependence of Stokes shift (difference between the energies of absorption and luminescence bands), (2) the temperature dependence of PL spectra, and (3) the spectral distribution (luminescence energy dependence) of decay time constants of luminescence~\cite{makino11,makino12,sun2,makino13}.
A typical example of the spectral distribution of decay time constants is shown
in the lowest curve of Fig.~\ref{fig:rawdata}(b).
Here, the temperature dependence of the PL spectrum in a QW in the case of magnesium composition of 0.27 is described in detail.

\begin{figure}[htb]
\includegraphics[width=.5\textwidth]{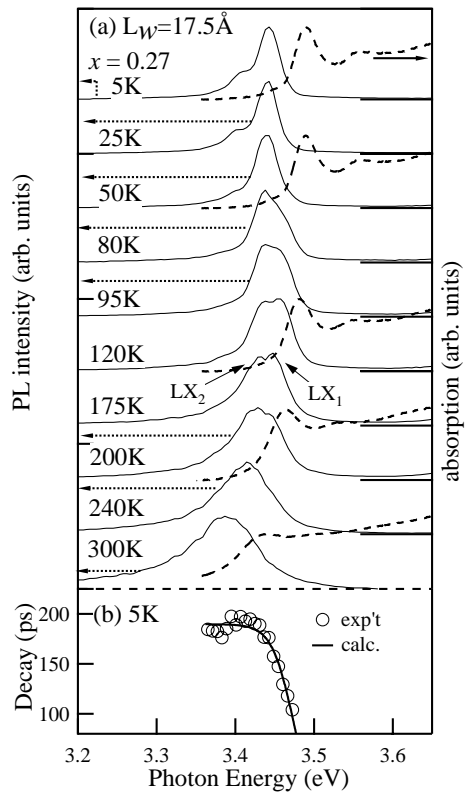}
\caption{(a) PL (solid line) and absorption (broken line) spectra in a ZnO
(17.5~\AA)/Mg$_{0.27}$Zn$_{0.73}$O MQW over the temperature range of 5 to
300~K. All of the spectra have been normalized and shifted in the vertical
direction for clarity. (b) PL decay times as a function of monitored photon
energy at 5~K in the same MQW. The dotted curve is results of the theoretical calculation based on the model of the excitonic localization.}
\label{spectrum-tv}
\end{figure}
\begin{figure}
\includegraphics[width=.5\textwidth]{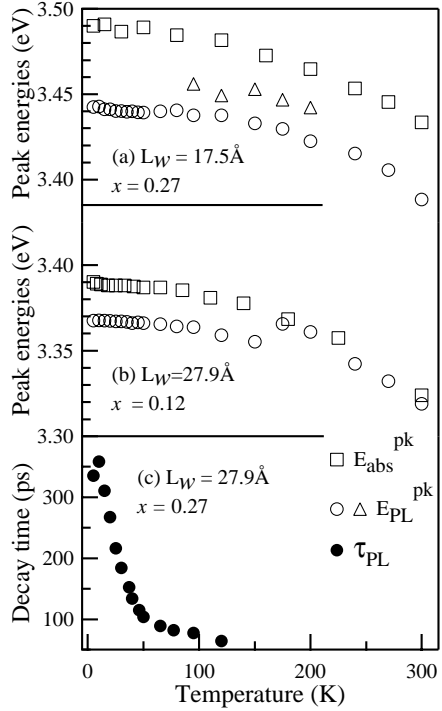}
\caption{PL (solid circles and triangles) and absorption (solid squares)
peak positions as a function of temperature in ZnO(17.5~\AA)/Mg$_{0.27}
$Zn$_{0.73}$O (a) and ZnO(27.9~\AA)/Mg$_{0.12}$Zn$_{0.88}$O (b) MQWs.
(c) Temperature dependence of PL decay times, $\tau_{PL}$, in
ZnO(27.9~\AA)/Mg$_{0.27} $Zn$_{0.73}$O MQWs at temperatures of 5--120~K~\protect\cite{makino11}.}
\label{fig:pkplot}
\end{figure}

Figure~\ref{spectrum-tv}(a) shows the temperature dependence of PL (solid line) and that of
absorption (broken line) spectra in ZnO(17.5~\AA)/Mg$_{0.27}$Zn$_{0.73}$O
MQWs over a temperature range of 5 to 300~K. It should be noted that
spectra obtained at temperatures between
95 and 200~K had two peaks, both of which originated from
a recombination of localized excitons. The separation of these peaks was
12 to 20~meV. Figure~\ref{spectrum-tv}(b) shows PL decay time as a function of monitored photon
energy at 5~K in the same MQW. The dotted curve is the results of theoretical calculation based on the model of the excitonic localization~\protect\cite{makino11,gourdon1}. Figure~\ref{fig:pkplot}(a) summarizes peak energies of the PL spectra ($E_{PL}^{\textrm{pk}}$) (solid circles and triangles) and the
excitonic absorption energy (solid squares) as functions of temperatures. It should be noted that the higher PL peak position does
not coincide with that of absorption spectra even at temperatures near room temperature. We also examined, for comparison, the temperature dependence of PL peak
energy in an MQW having a lower barrier height: a ZnO/Mg$_{0.12}$Zn$_{0.88}$O
MQW with a well width of 27.9~\AA. Figure~\ref{fig:pkplot}(b) shows the peak
energies of PL (circles) and absorption (squares) spectra in this sample.
In this case, the energies of luminescence and absorption are the almost same with
each other at temperatures near room temperature.
Two kinds of MQWs having different barrier
heights showed significantly
different temperature dependences of PL spectra.

Following a temperature rise, the PL energy of ZnO(17.5~\AA)/Mg$_{0.27}$Zn$_{0.73}$O
MQWs exhibited low energy shifts between 5K and 50K, blue-shifts between 50 and 200~K, and again shifts to a low energy side at temperatures higher than 200~K.
Furthermore, at temperatures between 95 and 200~K, the spectra had two peaks, both of which originated from
a recombination of localized excitons.
The absorption peak energies both in ZnO epilayers and in MQWs are
monotonically decreasing functions of temperature as was revealed in
previous studies~\cite{makino8,sun2}. This is attributed to the temperature-induced shrinkage of the fundamental energy gap.

In general, when a dominant PL
peak is assigned to a radiative recombination of localized excitons, its peak
energy blue-shifts with increase in temperature in a range of low temperatures and
red-shifts at higher temperatures. The $E_{PL}^{\textrm{pk}}$ blue-shifts and
continuously connects to that of free excitons due to thermal activation of
localized excitons. The $E_{PL}^{\textrm{pk}}$ of the free-excitonic emission
is a monotonically decreasing function of temperature due to the band gap
shrinkage. The temperature dependence shown in Fig.~\ref{fig:rawdata}(a) is,
however, different from the abovementioned typical behavior.
The temperature dependence of the recombination mechanism for localized
excitons is thought to be closely related to the temperature variation
in the decay time constant of their PL. Thus temperature dependence of PL decay times
($\tau_{PL}$) in an MQW with a $L_w$ of
27.9~\textrm{\AA} is shown in Fig.~7(c). The $\tau_{PL}$ values
exhibit nonmonotonical behavior with respect to temperature; the value increased in a low temperature range and decreased above a certain critical temperature.

The temperature dependence of the recombination mechanism for localized
excitons can be explained~\cite{yhcho2} as follows. For 5~K
\textless $T $\textless 50~K, the relatively long relaxation time of
excitons gives the excitons more opportunity to relax down into lower
energy tail states caused by the inhomogeneous potential fluctuations
before recombining. This is because radiative recombination
processes dominate in this
temperature range. This behavior produces a red-shift in the peak energy
position with increasing temperature. For 50~K \textless $T
$\textless 95~K, the exciton lifetimes decrease with increasing
temperature. Thus, these excitons recombine before reaching the lower
energy tail states. This behavior enhances a broadening of the higher-energy
side emission and leads to a blue shift in the peak energy. (iii) For 95~K
\textless $T $\textless 200~K, further enhancement of high-energy emission
components produces a new peak, as seen in Fig.~\ref{fig:pkplot}(a)
(triangles). (iv) At temperatures above 200~K, since the
excitons are less affected by the temperature-induced rapid change in their
lifetimes and since the relaxation rate of the excitons increased due to an increase in the phonon
population, blue-shift behavior becomes less pronounced. Since
the energy of a blue-shift is less than the temperature-induced band gap
shrinkage, the peak position again exhibits a red-shift behavior. As
mentioned above, the features of excitonic spontaneous emission in the
well layers are sensitively affected by the dynamics of
recombination of localized exciton states, which significantly vary with temperature. Readers should refer to the original articles
for
details~\cite{makino12}.
\section{Mechanism of stimulated emission in multiple quantum wells}
\label{Sect:7}
\begin{figure}
\includegraphics[width=.5\textwidth]{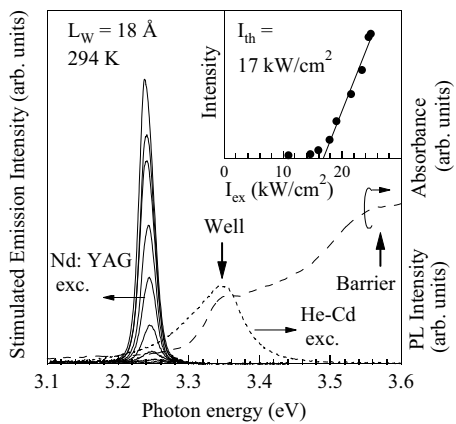}
\caption{Excitation intensity ($I_{ex}$) dependence of the stimulated
emission spectra obtained from a ZnO/Mg$_{0.12}$Zn$_{0.88}$O SL ($L_{w}$=1.8~nm) under the condition of
pulsed excitation at RT. Spontaneous PL (dotted line) under the condition of
continuous-wave (cw)
excitation and absorption (broken line) spectra are also shown. Inset
shows the integrated intensity of the stimulated emission peak as a
function of $I_{ex}$. Threshold intensity ($I_{th}$) is 17~kW/cm$^{2}$~\protect\cite{ohtomo8}.}
\label{stim_sl}
\end{figure}
As shown in Fig.~\ref{stim_sl}, the energies of spontaneous PL and absorption peaks are
almost the same at room-temperature~\cite{makino11}. The spontaneous emission spectrum was obtained under the condition of excitation using a 5-mW-power helium
cadmium laser operated in the continuous-wave mode, while the stimulated emission
spectrum was obtained under the condition of high-power excitation using a frequency-tripled mode-locked Nd:YAG laser (355~nm, 10~Hz, 15~ps) operated in the pulsed mode.
The power of excitation was varied, as is described later. The
agreement between the spontaneous emission and absorption peaks is an indication
of the well-regulated heterointerfaces as well as the small compositional
fluctuations in the barrier layers (well-depth fluctuations). Such room-temperature PL in the MQWs grown on the sapphire substrates was not found.

We performed high-power excitation experiments to determine the characteristics of stimulated emission in ZnO MQWs in the optical pumping
mode.
Figure~\ref{stim_sl} shows stimulated emission spectra of MQWs with $x=0.12$
and $L_{w}=17.5$~\textrm{$\AA$} that were obtained at room
temperature. Strong and sharp emission peaks were observed at 3.24~eV above
a very low threshold ($I_{th}$=17 kW/cm$^{2}$), and their integrated
intensities rapidly increased as the excitation intensity ($I_{ex}$) increased, as can be seen in the inset. Although such stimulated emission was not observed even at 4.2~K in the case using a sapphire substrate, RT stimulated emission in MQWs studied here was observed.
This is one of the significant improvements achieved by applying lattice-matching conditions to a substrate.

\begin{figure}
\includegraphics[width=.5\textwidth]{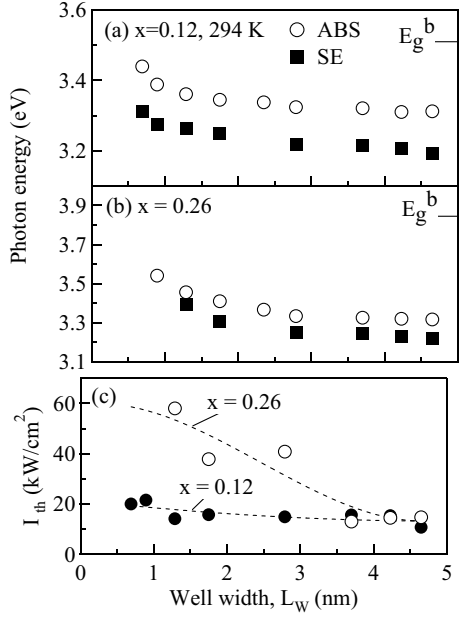}
\caption{Optical transition energies of subband absorption (open circles) and stimulated emission (closed squares) as a function of well layer thickness ($L_{w}$) for ZnO/Mg$_{x}$Zn$_{1-x}$O MQWs with $x=$0.12 (a) and $x=$0.26 (b). Band gap energy of the barrier layers ($E_{g}^{\rm b} $) is also shown. (c) $L_{w}$ dependence of the stimulated emission threshold ($I_{th}$) in MQWs with $x=$0.12 (closed circles) and $x=$0.26 (open circles). Stimulated emission did not occur for the $x=$0.26 films with $L_{w}$ below 1~nm since the excitation energy is lower than the absorption energy~\protect\cite{ohtomo8}.}
	\label{stim_well}
\end{figure}
The $L_w$ dependences of peak energies of the absorption and stimulated
emissions are summarized in Fig.~\ref{stim_well}. Similar dependence of the threshold is also
shown in Fig.~\ref{stim_well}(c). The stimulated emission energy is 100~meV lower than that
of the absorption peak. The lowest threshold value was 11 kW/cm$^{2}$ in the
case of $L_w$ of 47~\textrm{\AA}.

\begin{figure}
\includegraphics[width=.5\textwidth]{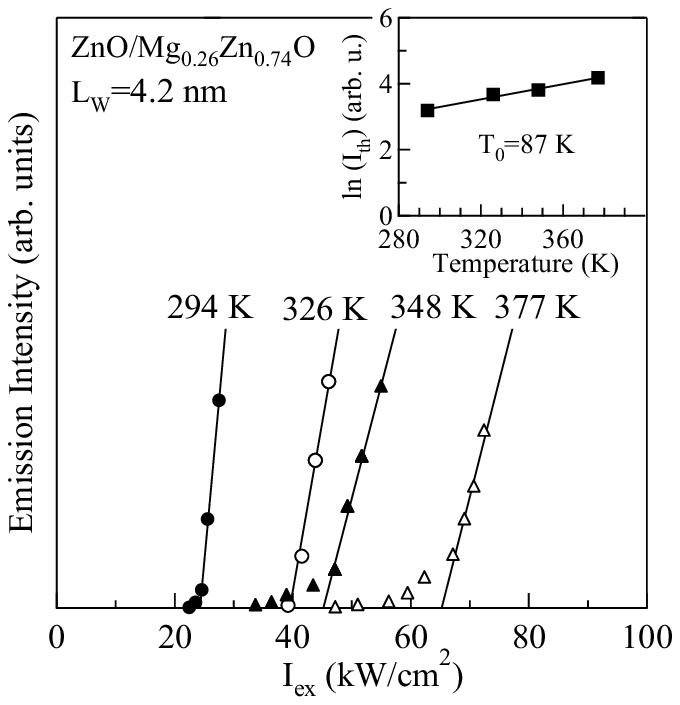}
\caption{Temperature dependence of emission intensity as a function of excitation intensity ($I_{ex}$) in a ZnO/Mg$_{0.26}$Zn$_{0.74}$O SL ($L_{W}=$4.2~nm). Inset shows the threshold intensity ($I_{th}$) as a function of temperature on a logarithmic scale~\protect\cite{ohtomo8}.}
\label{temp}
\end{figure}
We tested
the high-temperature operation of the stimulated emission from the viewpoint
of possible applications to devices. Figure~\ref{temp} shows the temperature dependence of the $I_{\rm stim}-I_{\rm ex}$
curves of a MQW with $x=$0.26 and $L_{w}=$4.2~nm in the temperature range
of 294~K to 377~K. Here, $I_{\rm stim} $ is the intensity of the
stimulated emission and $I_{\textrm{ex}}$ is the excitation intensity.
The threshold of the stimulated emission ($I_{th} $) increased gradually
with increasing temperature. The inset shows the temperature dependence of
$I_{th}$ on a logarithmic scale. Characteristic temperature, which is an index of stability of threshold characteristics with respect to temperature rise, was estimated to be 87~K~\cite{ohtomo8}. This was significantly higher than that of
a 55-nm-thick ZnO/sapphire (67~K)~\cite{ohtomophd}, which showed
excitonic laser action with a threshold of 24~kW/cm$^{2}$. We speculate that
this kind of improvement can be explained by the enhanced binding energy of
excitons due to the quantum confinement effect~\cite{sun1}.

In order to clarify the mechanism of stimulated emission of these MQWs, the temperature dependence of the stimulated-emission spectrum in the temperature range from low temperature to room temperature was estimated.
Figure 11 shows the temperature dependence of the peak energy of the stimulated emission band in the QWs (Mg concentration of 0.12, $L_w$s of 37.0\textrm{\AA } and 17.5\textrm{\AA }).
For comparison, the same plot for thin films of ZnO grown on sapphire substrates is shown in Fig.~11(a).
It is already clear that the RT stimulated emission band in these thin films is what is called the \textit{P}-line, which
one of the typical phenomena of high-density exciton effects.
Inelastic scattering between excitons (Auger-like process) gives rise to the appearance of this stimulated emission band.
One of the two excitons participating in the collisional events is ionized, whereas
the other is recombined radiatively after collision.
At sufficiently low temperatures, peak energy of the relevant stimulated emission band is lower than the resonance energy of an exciton, the energy difference of which is equal to the EBE.
The temperature dependence of the peak energy difference in QWs shows the same behavior as that of ZnO, as can be clearly understand in Fig.~11.
\begin{figure}
\includegraphics[width=.5\textwidth]{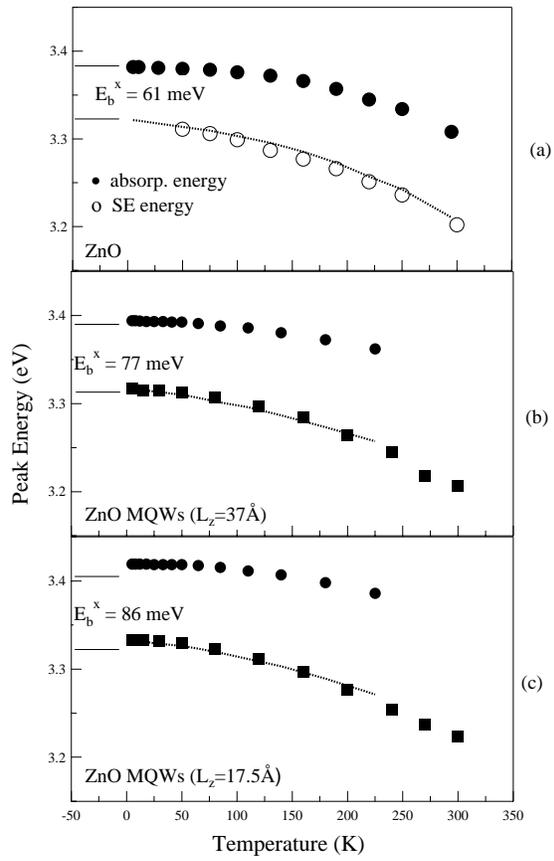}
\caption{Temperature dependence of peak energy of the \textit{P} band (open
circles) and free exciton energy (filled circles) in a ZnO epitaxial layer (a)
and in ZnO/Zn$_{0.88}$Mg$_{0.12} $O MQWs (b) and (c)~\protect\cite{sun1}.}
\label{peak_pos}
\end{figure}
Thus, as a result of careful comparison with that of a ZnO thin film, it became clear that the mechanism of this stimulated emission is inelastic scattering processes between the excitons. Therefore, the $L_w$ dependence of EBEs can be experimentally determined from analysis of the energy position of the \textit{P}-line.
As shown in Fig.~\ref{bindingenergy} (right axis), when $L_w$ decreased, the EBE increased up to about 90~meV and exceeded $\hbar \omega_{\rm LO} $ (72~meV). This enhancing effect is due to the quantum confinement effect. It has been revealed that these estimated EBEs approximately agree with
the theoretical values~\cite{coli1,sunfull1}. If QW structures were used, the phonon scattering process and thermo-broadening effect
of excitonic linewidth can be controlled. These are favorable from the viewpoint
of application.

Combinatorial concept-aided techniques adopted in the growth of our samples
suppressed the variations in crystal growth conditions and
hence the undesired uncertainty in the deduced spectroscopic
results.
In a related review article, the benefit of the combinatorial technique is described in detail~\cite{koinuma2}. For example, the well width dependence of
the radiative and nonradiative recombination times of localized excitons
was estimated by time-resolved photoluminescence spectroscopy~\cite{chia2,makino13}. Well width dependence of \textit{biexciton}
binding energy was also
estimated~\cite{sun3}. In addition, optical properties of (Cd,Zn)O/(Mg,Zn)O MQWs have not ever explored so far
except our work~\cite{makino13}. These structures are advantageous from the
viewpoint of almost perfect in-plane lattice-matching.

\section{Conclusions}
\label{Sect:8}
If a short-wavelength (UV region) semiconductor laser using exciton transitions in low-dimensional crystals (e.g., quantum wells) of ZnO can be produced, it
is expected that threshold of the laser action will become lower than that of a laser using an ordinary recombination mechanism of electrons and holes.
This study has shown, through experiments on optical excitation, that it is possible to produce a laser with a low threshold (i.e., small driving current density).
Laser action due to the effect of a longitudinal
cavity using grain boundaries is no longer observed in ZnO epitaxial
layers deposited on lattice-matched substrates, because their level of crystallinity has been greatly improved. Moreover, they no longer show a polycrystalline nature
(i.e., assembly of hexagonal pillar prisms). The corresponding resonance cavity structure has disappeared.

Moreover, a \textit{p-n} junction is essential for the production of a current injection laser. A ZnO crystal usually shows \textit{n}-type conductivity and it
has
not been possible to produce a \textit{p}-type layer with low resistance. Such a
feature of
widegap semiconductors is called unipolarity in carrier doping. In this
chapter, we have not discussed this problem.

Such high-quality ZnO QWs deposited on lattice-matched substrates are expected
to have many applications for UV optoelectronics devices.
Various applications of ZnO, such as its use in transparent electric-conduction films or a surface acoustic-wave device, are well-known.
New function of this substance have been found through research on high-quality single-crystalline thin films grown by the use of the laser-assisted molecular-beam-epitaxy.
New physical properties of other oxides may also be discovered in the future.

\section*{Acknowledgements}

The quantum wells used as the target of optical characterization in our study were produced by Dr.~A.~Ohtomo and K.~Tamura using equipment originally constructed by Dr.~Y.~Matsumoto, Tokyo Institute of Technology, Japan.
We would like to express our gratitude to the abovementioned researchers as well as Dr.~Ngyuen Tien Tuan, Dr.~H. D. Sun and C.~H. Chia, Institute of Physical and Chemical Research, Japan, for their assistance in all aspects of our research.

\newpage 

\newpage 
\begin{thebibliography}{10}

\bibitem{akasaki1}
Hiromitsu Sakai, Takashi Koide, Hiroyuki Suzuki, Machiko Yamaguchi, Shiro
  Yamasaki, Masayoshi Koike, Hiroshi Amano, and Isamu Akasaki.
\newblock {GaN/GaInN/GaN} double heterostructure light emitting diode
  fabricated using plasma-assisted molecular beam epitaxy.
\newblock {\em Jpn. J. Appl. Phys. Part 2}, 34(11A):L1429--L1431, November
  1995.

\bibitem{nakamura_spr}
S.~Nakamura, S.~Pearton, and G.~Fasol.
\newblock {\em The Blue Laser Diode}.
\newblock Springer-Verlag Berlin, 2000.

\bibitem{ohtomo1}
A.~Ohtomo, M.~Kawasaki, T.~Koida, K.~Masubuchi, H.~Koinuma, Y.~Sakurai,
  Y.~Yoshida, T.~Yasuda, and Y.~Segawa.
\newblock {Mg$_x$Zn$_{1-x}$O} as a {II-IV} widegap semiconductor alloy.
\newblock {\em Appl. Phys. Lett.}, 72(19):2466, May 1998.

\bibitem{kawasaki2}
M.~Kawasaki, A.~Ohtomo, R.~Shiroki, I.~Ohkubo, H.~Kimura, G.~Isoya, T.~Yasuda,
  Y.~Segawa, and H.~Koinuma.
\newblock {ZnO} quantum structures towards {UV} diode lasers.
\newblock In {\em Extended Abstracts of the 1998 International Conference on
  Solid State Devices and Materials}, page 356, Hiroshima, Japan, 1998.
  Business Ctr. Acad. Soc. Jpn.

\bibitem{makino14}
T.~Makino, Y.~Segawa, M.~Kawasaki, A.~Ohtomo, R.~Shiroki, K.~Tamura, T.~Yasuda,
  and H.~Koinuma.
\newblock Band gap engineering based on {(Mg,Zn)O} and {(Cd,Zn)O} ternary alloy
  films.
\newblock {\em Appl. Phys. Lett.}, 78(9):1237, Feb. 2001.

\bibitem{ohtomo4}
A.~Ohtomo, M.~Kawasaki, I.~Ohkubo, H.~Koinuma, T.~Yasuda, and Y.~Segawa.
\newblock Structure and optical properties of {ZnO/Mg$_{\rm 0.2}$Zn$_{\rm
  0.8}$O} superlattices.
\newblock {\em Appl. Phys. Lett.}, 75(1):980, 1999.

\bibitem{ohtomo7}
A.~Ohtomo, R.~Shiroki, I.~Ohkubo, H.~Koinuma, and M.~Kawasaki.
\newblock Thermal stability of supersaturated {Mg$_{x}$Zn$_{1-x}$O} alloy films
  and {Mg$_{x}$Zn$_{1-x}$O/ZnO} heterointerfaces.
\newblock {\em Appl. Phys. Lett.}, 75(26):4088, Dec. 1999.

\bibitem{makino17}
T.~Makino, Y.~Segawa, A.~Ohtomo, K.~Tamura, T.~Yasuda, M.~Kawasaki, and
  H.~Koinuma.
\newblock Strain effect on exciton resonance energies of zno epitaxial layers.
\newblock {\em Appl. Phys. Lett.}, 79(9):1282, August 2001.

\bibitem{kimizuka1}
N.~Kimizuka and T.~Mohri.
\newblock {\em J. Solid State Chem.}, 78:98, 1989.

\bibitem{ohtomo5}
A.~Ohtomo, K.~Tamura, K.~Saikusa, T.~Takahashi, T.~Makino, Y.~Segawa,
  H.~Koinuma, and M.~Kawasaki.
\newblock Single crystalline {ZnO} films grown on lattice matched
  {ScAlMgO$_{4}$(0001)} substrates.
\newblock {\em Appl. Phys. Lett.}, 75(17):2635, 1999.

\bibitem{yfchen3}
Several efforts to grow not MQWs but epilayers are currently being made using
  MgO or MgAl$_{2}$O$_{4}$(111) substrates by means of plasma-assisted MBE,
  whose misfits are smaller than that of sapphire (Y. F. Chen {et~al\/}).

\bibitem{matsumoto1}
Y.~Matsumoto, M.~Murakami, Z.~W. Jin, A.~Ohtomo, M.~Lippmaa, M.~Kawasaki, and
  H.~Koinuma.
\newblock Combinatorial laser molucular epitaxy ({MBE}) growth of {Mg-Zn-O}
  alloy for band gap engineering.
\newblock {\em Jpn. J. Appl. Phys., Part2}, 38(2(6A/B)):L603, Jun. 1999.

\bibitem{koinuma2}
H.~Koinuma, H.~N. Aiyer, and Y.~Matsumoto.
\newblock Combinatorial solid state materials science and technology.
\newblock {\em Sci. \& Tech. Adv. Mater.}, 1(1):1, 2000.

\bibitem{ohtomo9}
A.~Ohtomo, T.~Makino, K.~Tamura, Y.~Matsumoto, Y.~Segawa, Z.~K. Tang, G.~K.~L.
  Wang, H.~Koinuma, and M.~Kawasaki.
\newblock High throughput optimizations of alloy and doped films based on {ZnO}
  and parallel synthesis of {ZnO/(Mg,Zn)O} quantum wells using combinatrial
  laser {MBE} towards ultraviolet laser.
\newblock {\em Proceedings of SPIE}, 3941:70, 2000.

\bibitem{makino9}
T.~Makino, G.~Isoya, Y.~Segawa, C.~H. Chia, T.~Yasuda, M.~Kawasaki, A.~Ohtomo,
  K.~Tamura, Y.~Matsumoto, and H.~Koinuma.
\newblock Optical characterization for combinatrial system based on
  semiconductor {ZnO}.
\newblock In G.~Jabbour, editor, {\em Proceedings of the 1st International
  Conference on Combinatorial and Composition Spread Techniques in Material and
  Device Development, San Jose}, volume 3941, page~28, Bellingham, 2000. SPIE.

\bibitem{ohtani1}
M.~Ohtani, T.~Fukumura, M.~Kawasaki, K.~Omote, T.~Kikuchi, J.~Harada,
  A.~Ohtomo, M.~Lippmaa, T.~Ohnishi, D.~Komiyama, R.~Takahashi, Y.~Matsumoto,
  and H.~Koinuma.
\newblock Concurrent x-ray diffractometer for high throughput structural
  diagnosis of epitaxial thin films.
\newblock {\em Appl. Phys. Lett.}, 79(22):3594, Nov. 2001.

\bibitem{ohtani2}
M.~Ohtani, T.~Fukumura, M.~Kawasaki, K.~Omote, T.~Kikuchi, J.~Harada, and
  H.~Koinuma.
\newblock Concurrent evaluation of strain in heteroepitaxial thin films with
  continuous lattice mismatch spread.
\newblock {\em Appl. Phys. Lett.}, 80(12):2066, Mar. 2002.

\bibitem{makino19}
T.~Makino, K.~Tamura, C.~H. Chia, Y.~Segawa, M.~Kawasaki, A.~Ohtomo, and
  H.~Koinuma.
\newblock Optical properties of zno:al epilayers: Observation of
  room-temperature many-body absorption-edge singularity.
\newblock {\em Phys. Rev. B}, 65(12):121201(R), March 2002.

\bibitem{LBZincoxide}
E.~Mollwo.
\newblock In O.~Madelung, M.~Schulz, and H.~Weiss, editors, {\em
  Semiconductors: Physics of II-VI and I-VII Compounds, Semimagnetic
  Semiconductors}, volume~17 of {\em Landolt-B{\"{o}}rnstein New Series},
  page~35. Springer, Berlin, 1982.

\bibitem{liang1}
W.~Y. Liang and A.~D. Yoffe.
\newblock Transmission spectra of {ZnO} single crystals.
\newblock {\em Phys. Rev. Lett.}, 20(2):59, 1968.

\bibitem{makino8}
T.~Makino, C.~H. Chia, N.~T. Tuan, Y.~Segawa, M.~Kawasaki, A.~Ohtomo,
  K.~Tamura, and H.~Koinuma.
\newblock Exciton spectra in {ZnO} epitaxial layers on lattice-matched
  substrates grown with laser-molecular-beam epitaxy.
\newblock {\em Appl. Phys. Lett.}, 76(24):3549, Jun. 2000.

\bibitem{ohtomophd}
Akira Ohtomo.
\newblock {\em Quantum structures and ultraviolet light-emitting devices based
  on {ZnO} thin films grown by laser molecular-beam epitaxy}.
\newblock PhD thesis, Tokyo Institute of Technology, Mar. 2000.

\bibitem{ohtomo2}
A.~Ohtomo, M.~Kawasaki, Y.~Sakurai, I.~Ohkubo, R.~Shiroki, Y.~Yoshida,
  Y.~Sakurai, T.~Yasuda, Y.~Segawa, and H.~Koinuma.
\newblock Fabrication of alloys and superlattices based on zno towards
  ultraviolet laser.
\newblock {\em Mat. Sci. Eng. B}, 56(2):263, Nov. 1998.

\bibitem{makino11}
T.~Makino, N.~T. Tuan, H.~D. Sun, C.~H. Chia, Y.~Segawa, M.~Kawasaki,
  A.~Ohtomo, K.~Tamura, and H.~Koinuma.
\newblock Room-temperature luminescence of excitons in {ZnO/(Mg,Zn)O}
  multi-quantum wells on lattice-matched substrates.
\newblock {\em Appl. Phys. Lett.}, 77(7):975, 2000.

\bibitem{goldman}
I.~I. Gol'dman and V.~Krivchenokov.
\newblock {\em Problems of Quantum Mechanics}, page~60.
\newblock Addison-Wesley, Reading, Mass., 1961.

\bibitem{huemmer1}
K.~H{\"{u}}mmer.
\newblock Interband magnetoreflection in {ZnO} bulk.
\newblock {\em Phys. Status Solidi}, 56:249, 1973.

\bibitem{coli1}
Giuliano Coli and K.~K. Bajaj.
\newblock Excitonic transitions in zno/mgzno quantum well heterostructures.
\newblock {\em Appl. Phys. Lett.}, 78(19):2861, May 2001.

\bibitem{lautenschlager1}
P.~Lautenschlager, M.~Garriga, S.~Logothetidis, and M.~Cardona.
\newblock Interband critical points of {GaAs} and their temperature dependence.
\newblock {\em Phys. Rev. B}, 35:9174, 1987.

\bibitem{pelekanos1}
N.~T. Pelekanos, J.~Ding, M.~Hagerott, A.~V. Nurmikko, H.~Luo, N.~Samarth, and
  J.~K. Furdyna.
\newblock Quasi-2-dimensional {(Zn,Cd)Se/ZnSe} quantum wells reduced exciton
  lo-phonon coupling due to confinement effects.
\newblock {\em Phys. Rev. B}, 45:6037, 1992.

\bibitem{sun2}
H.~D. Sun, T.~Makino, N.~T. Tuan, Y.~Segawa, M.~Kawasaki, A.~Ohtomo, K.~Tamura,
  and H.~Koinuma.
\newblock Temperature dependence of the exciton linewidth in {ZnO/(Mg,Zn)O}
  multi-quantum wells.
\newblock {\em Appl. Phys. Lett.}, 78(17):2464, 2001.

\bibitem{sunfull1}
H.~D. Sun, T.~Makino, Y.~Segawa, M.~Kawasaki, A.~Ohtomo, K.~Tamura, and
  H.~Koinuma.
\newblock Enhancement of exciton binding energies in zno/znmgo multiquantum
  wells.
\newblock {\em J. Appl. Phys.}, 91(4):1993, Feb. 2002.

\bibitem{makino12}
T.~Makino, N.~T. Tuan, H.~D. Sun, C.~H. Chia, Y.~Segawa, M.~Kawasaki,
  A.~Ohtomo, K.~Tamura, M.~Baba, H.~Akiyama, T.~Suemoto, S.~Saito, T.~Tomita,
  and H.~Koinuma.
\newblock Temperature dependence of near ultraviolet photoluminescence in
  {ZnO/(Mg,Zn)O} multi-quantum wells.
\newblock {\em Appl. Phys. Lett.}, 78(14):1979, 2001.

\bibitem{makino13}
T.~Makino, N.~T. Tuan, Y.~Segawa, C.~H. Chia, M.~Kawasaki, A.~Ohtomo,
  K.~Tamura, and H.~Koinuma.
\newblock Radiative and nonradiative recombination processes in lattice-matched
  {(Cd,Zn)O/(Mg,Zn)O} multiquantum wells.
\newblock {\em Appl. Phys. Lett.}, 77(11):1632, 2000.

\bibitem{gourdon1}
C.~Gourdon and P.~Lavallard.
\newblock Exciton transfer between localized states in {CdSSe} alloys.
\newblock {\em Phys. Status Solidi (b)}, 153:641, 1989.

\bibitem{yhcho2}
C.~Yong-Hoon, B.~D. Little, G.~H. Gainer, J.~J. Song, S.~Keller, U.~K. Mishra,
  and S.~P. DenBaars.
\newblock Carrier dynamics of abnormal temperature-dependent emission shift in
  mocvd-grown ingan epilayers and ingan/gan quantum wells.
\newblock {\em MRS Int. J. Nitride. Res.}, 4S1:G2.4, 1999.

\bibitem{ohtomo8}
A.~Ohtomo, K.~Tamura, M.~Kawasaki, T.~Makino, Y.~Segawa, Z.~K. Tang, G.K.L.
  Wong, Y.~Matsumoto, and H.~Koinuma.
\newblock Room-temperature stimulated emission of excitons in {ZnO/(Mg,Zn)O}
  superlattices.
\newblock {\em Appl. Phys. Lett.}, 77(14):2204, Oct. 2000.

\bibitem{sun1}
H.~D. Sun, T.~Makino, N.~T. Tuan, Y.~Segawa, Z.~K. Tang, G.~K.~L. Wong,
  M.~Kawasaki, A.~Ohtomo, K.~Tamura, and H.~Koinuma.
\newblock Stimulated emission induced by the inelastic exciton-exciton
  scattering in {ZnO/(Mg,Zn)O} multi-quantum wells.
\newblock {\em Appl. Phys. Lett.}, 77(26):4250, 2000.

\bibitem{chia2}
C.~H. Chia, T.~Makino, Y.~Segawa, M.~Kawasaki, A.~Ohtomo, K.~Tamura, and
  H.~Koinuma.
\newblock Well-width dependence of radiative and nonradiative recombination
  times in zno/mgzno multiple quantum wells.
\newblock {\em J. Appl. Phys.}, 90(7):3650, Nov. 2001.

\bibitem{sun3}
H.~D. Sun, T.~Makino, Y.~Segawa, M.~Kawasaki, A.~Ohtomo, K.~Tamura, and
  H.~Koinuma.
\newblock Biexciton emission from {ZnO/(Mg,Zn)O} multi-quantum wells.
\newblock {\em Appl. Phys. Lett.}, 78(22):3385, May 2001.

\end{thebibliography}
\end{document}